\documentclass[conference]{IEEEtran}
\IEEEoverridecommandlockouts

\usepackage{cite}
\usepackage{amsmath,amssymb,amsfonts}
\usepackage{algorithmic}
\usepackage{graphicx}
\usepackage{textcomp}
\usepackage{xcolor}
\usepackage{multirow}

\def\BibTeX{{\rm B\kern-.05em{\sc i\kern-.025em b}\kern-.08em
    T\kern-.1667em\lower.7ex\hbox{E}\kern-.125emX}}
\begin{document}

\title{A Mamba-based Network for Semi-supervised 
Singing Melody Extraction Using Confidence Binary Regularization\\

}
\author{\IEEEauthorblockN{Xiaoliang He}
\IEEEauthorblockA{\textit{School of Comp. Science \& Technology 
} \\
\textit{Donghua University}\\
Shanghai, China\\
}
\and
\IEEEauthorblockN{Kangjie Dong}
\IEEEauthorblockA{\textit{School of Comp. Science \& Technology 
} \\
\textit{Donghua University}\\
Shanghai, China\\
}
\and
\IEEEauthorblockN{Jingkai Cao}
\IEEEauthorblockA{\textit{School of Comp. Science \& Technology 
} \\
\textit{Donghua University}\\
Shanghai, China\\
}

\and
\IEEEauthorblockN{Shuai Yu*\thanks{*Shuai Yu and Wei Li are corresponding authors.}}
\IEEEauthorblockA{\textit{School of Comp. Science \& Technology 
} \\
\textit{Donghua University}\\
Shanghai, China\\
}

\and
\IEEEauthorblockN{Wei Li*}
\IEEEauthorblockA{\textit{
School of Comp. Science \& Technology
} \\
\textit{Fudan University}\\
Shanghai, China\\
}
\and
\IEEEauthorblockN{Yi Yu}
\IEEEauthorblockA{\textit{
Grad. School of Adv. Science \& Eng.
} \\
\textit{Hiroshima University}\\
Hiroshima, Japan\\
}

}
\maketitle

\begin{abstract}
Singing melody extraction (SME) is a key task in the field of music information 
retrieval.
However, existing 
methods are facing several limitations: firstly, 
prior models use
transformers
to capture the contextual dependencies, 
which requires quadratic computation resulting in low efficiency in the inference stage.
Secondly, prior works typically rely on frequency-supervised methods to estimate the 
fundamental frequency (f0), 
which ignores that the
musical performance is actually based on notes. 
Thirdly, transformers typically require large amounts of labeled data 
to achieve optimal performances, but the SME task lacks of sufficient
annotated data.
To address these issues, in this paper, we propose a 
mamba-based network, called SpectMamba, for semi-supervised singing melody extraction 
using confidence binary regularization.
In particular, we begin by introducing vision mamba to achieve 
computational
linear complexity. Then, we propose a novel note-f0 decoder that
allows the model to better mimic the musical performance. 
Further, to alleviate the scarcity of the labeled data, we introduce a
confidence binary regularization (CBR) module to leverage the unlabeled data
by maximizing the probability of the correct classes.
The proposed method is evaluated on several public datasets
and the conducted experiments demonstrate the effectiveness of our 
proposed method \footnote{Codes are available in https://github.com/Tinkle01/SpectMamba}.

\end{abstract}

\begin{IEEEkeywords}
Singing Melody Extraction, Semi-Supervised, Mamba, Music Information Retrieval.
\end{IEEEkeywords}

\section{Introduction}
Singing melody extraction aims to extract the fundamental frequency contour from 
polyphonic music \cite{b30,b31,b32}. 
As a fundamental task in the field of music information retrieval (MIR),
it has become the cornerstone of many downstream scenarios, such as song identiﬁcation \cite{b1}, 
query-by-humming \cite{b3}, 
voice separation \cite{b4}, and music recommendation \cite{b5}. 

Recently, transformer-based models \cite{b17,b18,b19}
have achieved remarkable successes in SME. 
However, existing
methods are facing several limitations:
firstly, transformers require quadratic computation resulting in low efficiency in
the inference stage.
Secondly, prior works \cite{b26, b27, b28, b29}
typically rely on frequency-supervised methods to estimate the 
fundamental frequency.
Little research has studied the role of notes in improving the f0 estimation. 
Chen et al. \cite{b19} 
employed tone and octave supervision in SME and 
achieved outstanding‌ performance.
However, processing tone and octave separately makes the model hardly simulate musical performance,
as this logic is actually based on notes.
Furthermore, existing melody extraction models \cite{a1, a2, a3, a4, a5} 
mostly rely on the annotated music data,
and the lack of sufficient labeled data limits further improvements.

To address the above mentioned problems, 
we propose a mamba-based network for semi-supervised singing melody extraction. 
Specifically, inspired by vision mamba \cite{b7}, we segment the spectrogram into temporal 
patch sequences and process the melody representation with bidirectional state 
space models. 
The linear complexity makes the proposed model more efficient than transformers.
Moreover, we propose to use a coarse-to-fine method
to simulate the characteristics of musical performance.
We first estimate the note sequence
to
capture the overall contour and key transitions in the music using a coarse-grained approach. 
This step provides valuable prior knowledge for the subsequent fine-grained f0 estimation. 
Given that each note corresponds to a specific frequency range, we then perform fine-grained 
f0 estimation by leveraging the prior knowledge from the predicted notes.
Through such a coarse-to-fine manner, the proposed method can obtain 
more accurate results.

To deal with the scarcity of 
labeled data, inspired by AllMatch \cite{c30},
we use confidence binary regularization and
propose a semi-supervised SME module. The CBR module splits the 
prediction of each unlabeled sample into two parts: top-k predictions as its positive 
part and the rest as the negative part.
The strategy introduces consistency supervision 
for unlabeled data by encouraging consistent 
classifications across weakly and strongly augmented 
versions of the same sample.

Three technical contributions are made in this work: i) we introduce a mamba-based network, 
which achieves high efficiency in the inference stage; 
ii) the proposed note-f0 decoder utilizes a coarse-to-fine method to
simulate the characteristics of musical performance;
iii) we introduce a CBR module to leverage the
unlabeled data by maximizing the probability of the correct
class.
\begin{figure}[t]
    \centerline{\includegraphics[scale=0.35]{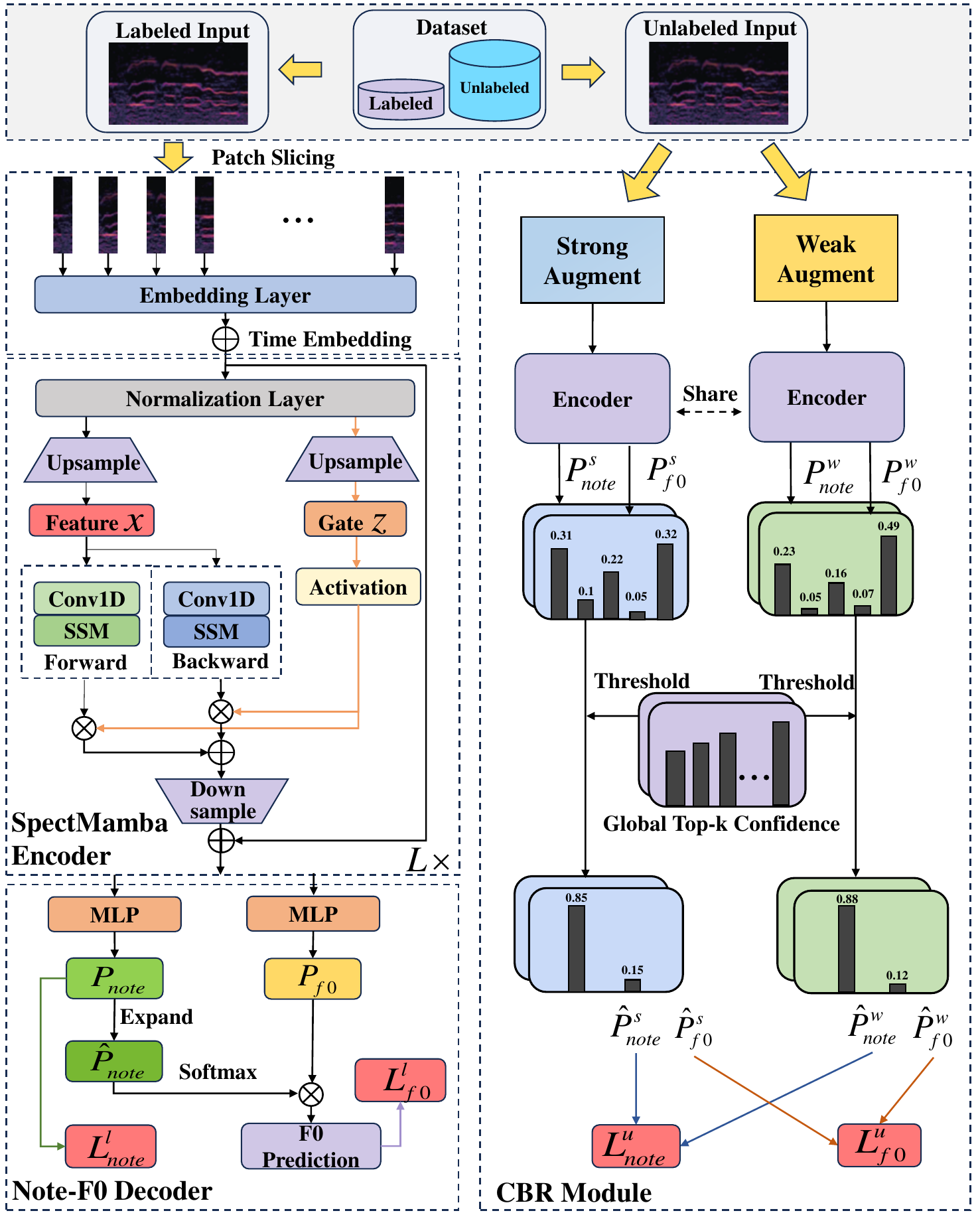}}
    \caption{The framework of the proposed SpectMamba.}
    \label{fig}
    \end{figure}

\section{PROPOSED MODEL}
The overall architecture of our proposed model is illustrated in Fig. 1. 
In this section, we will introduce 
the SpectMamba encoder, the note-f0 decoder and the CBR module, respectively.

\subsection{Semi-supervised Learning Setup}
In this paper, the inputs are from both labeled and unlabeled 
data. For the input data, the music signal can be denoted 
as $D = \{ {D_l},{D_u}\}$. ${D_l} = \{ ({x_1},{y_1}),({x_2},{y_2}),...,({x_M},{y_M})\}$ 
and ${D_u} = \{ {u_1},{u_2},...,{u_N}\} $ denote the labeled music data and 
unlabeled music data, respectively. $M$ and $N$ are the number of 
labeled and unlabeled data.
The learning objective function is constructed in the following form:
\begin{equation}
    \mathop {\min }\limits_\theta  \{ {L_l}({D_l},\theta ) + \omega {L_u}({D_u},\theta )\},
    \label{eq2} 
    \end{equation}
    where $L_l$ is the loss function of the supervised learning 
    and $L_u$ is the loss function of the unsupervised learning.
    $\omega$ is a non-negative parameter, $\theta$ represents the parameters
    of our proposed framework.
\subsection{SpectMamba Encoder}\label{AA}
To handle spectrograms, 
we slice the original spectrogram $S \in {\mathbb{R}^{C \times F \times T}}$ along the time 
dimension, where $C$, $F$, $T$ denote the number of channels, frequency bins 
and time steps, respectively,
resulting in a patch token sequence $P = \{ {P_1},{P_2},...,{P_T}\}$. Then, 
through an embedding layer, 
each patch token is projected 
as a token embedding. Next, we add the time embedding ${E_{pos}}$ into obtained
token embedding to obtain the input token sequence $I = \{ {I_1},{I_2},...,{I_T}\} $
of SpectMamba encoder:
\begin{equation}
{I} = Embedding({P}) + {E_{pos}}.
\label{eq1} 
\end{equation}
Then the sequence $I$ is fed into the SpectMamba encoder to 
extract task-specific features. Specifically, 
the SpectMamba encoder has L layers, the input of the $l$-th layer can be denoted as
$I^l$. $I^l$ is first normalized and then upsampled through a linear layer 
to obtain the feature $\mathbf{x}$ and gate $\mathbf{z}$. $\mathbf{x}$ is then 
processed from both forward and backward directions. For the forward direction,
following Vim \cite{b7}, 
we apply a 1-D convolution to $\mathbf{x}$ and feed it
into State Space Model (SSM) to obtain the forward hidden feature $\boldsymbol{y}_{forw}$. For the backward direction, 
the 1-D convolution and SSM are applied in the reverse order and then 
restored to the original order before outputing backward hidden feature $\boldsymbol{y}_{back}$:
\begin{equation}
{\boldsymbol{y}_{forw}} = SS{M^{forw}}(Con{v^{forw}}(MLP(Norm({I^l})))),
\label{2} 
\end{equation}
\begin{equation}
{\boldsymbol{y}_{back}} = SS{M^{back}}(Con{v^{back}}(MLP(Norm({I^l})))).
\label{3} 
\end{equation}
Finally, the $\boldsymbol{y}_{forw}$ and $\boldsymbol{y}_{back}$
are computed by the gate $\mathbf{z}$ and added to the residual ${I^l}$ to 
obtain the next layer input $I^{l + 1}$,
\begin{equation}
\boldsymbol{y}{'_{forw}} = {\boldsymbol{y}_{forw}} \odot SiLU(\mathbf{z}),
\label{4} 
\end{equation}
\begin{equation}
\boldsymbol{y}{'_{back}} = {\boldsymbol{y}_{back}} \odot SiLU(\mathbf{z}),
\label{5} 
\end{equation}
\begin{equation}
    {I^{l + 1}} = MLP(\boldsymbol{y}{'_{forw}} + \boldsymbol{y}{'_{back}}) + {I^l},
    \label{6} 
    \end{equation}
where $SiLU$ \cite{silu} serves as the activation function and gate $\mathbf{z}$
regulates the contribution of $\boldsymbol{y}_{forw}$ and $\boldsymbol{y}_{back}$ 
to the next layer.

\subsection{Note-F0 Decoder}
The note-f0 decoder consists of two components: the f0 prediction part and 
the note prediction part. Firstly, the note-f0 decoder takes the
output of the $L$-layer encoder ${I^{l + 1}}$
as the input. Then, the decoder generates f0 prediction 
${P_{f0}} \in {\mathbb{R}^{B \times (F + 1) \times T}}$ and note 
prediction ${P_{note}} \in {\mathbb{R}^{B \times (V + 1) \times T}}$ 
through two separate MLP layers. Here, $B$ denotes the batch size, $F$ denotes the number of f0 
classes and $V$ denotes the number of note classes. We use an extra one-dimensional feature 
as non-melody detection frame \cite{b14}. 
In order to enable note information to guide f0 prediction, we first
apply a repetition method to expand ${P_{note}}$ 
to ${\hat P_{note}} \in {\mathbb{R}^{B \times (F + 1) \times T}}$.
Since a single note corresponds to a range of frequencies, multiple f0 classes 
belong to the same note class. 
We then replicate note classes
to ensure that each f0 class is aligned with its corresponding note class. 
Next, a softmax function is applied to ${\hat P_{note}}$ to generate attention weights,
indicating the importance of different f0 classes. Finally, 
the refined melody prediction ${\hat P_{f0}}$ is obtained through a MLP layer. 
Cross-entropy loss ($CE$) is employed on the labeled data 
to calculate both note and f0 losses:
\begin{equation}
    {\hat P_{f0}} = {\kern 1pt} MLP({\rm{Softmax}}({\hat P_{note}}) \cdot {P_{f0}}),
    \label{7} 
    \end{equation}
    \begin{equation}
        {L_l} = CE({\hat P_{f0}},{Q_{f0}}) + CE({P_{note}},{Q_{note}}),
        \label{8} 
        \end{equation}
where ${Q_{f0}}$ denotes the label of the f0, 
${Q_{note}}$ denotes the label of the note.

\subsection{Confidence Binary Regularization}
The aim of the CBR module is to encourage consistent
classifications between weakly and strongly augmented data by introducing 
consistency supervision for unlabeled data.
To achieve this, weak augmentation (e.g., noising) and strong 
augmentation (e.g., volume adjustment, reverb) are first applied separately 
to the unlabeled data. Both augmented versions are then fed into the model, 
which outputs four prediction maps: 
1) f0 prediction maps from weak augmentation, 
2) note prediction maps from weak augmentation, 
3) f0 prediction maps from strong augmentation and 
4) note prediction maps from strong augmentation.

Since the pipeline for calculating CBR loss for note and 
f0 is similar, we only introduce the f0 CBR 
loss calculation for brevity. 
The CBR module selects the top-k predictions from each unlabeled sample 
as positive classes with the rest as negative classes. 
Considering the varying difficulty 
of melody extraction across different samples, 
we use the exponential moving average (EMA) to dynamically update global
top-$k$ confidence $\mu _t$. 
This adjustment accounts for both the previous global confidence $\mu _{t-1}$ 
and the current local confidence $p_t$, which is calculated as:
\begin{equation}
p_t = \frac{1}{B}\sum\limits_{i = 1}^B {\sum\limits_{j = 1}^k {{p_{i,{c_j}}}} },
\label{7} 
\end{equation}
where $c_j$ represents the $j$-th highest confidence class in the prediction 
map for the given sample $p_i$, $B$ represents batch size, $t$ represents the 
training step and $k$ represents the number of positive classes.
After determining $\mu _t$, 
we rank the f0 class predictions for each sample from highest to lowest. Then, 
we sum the top-$k$ predictions, adjusting 
$k$ until the cumulative value exceeds the global top-$k$ confidence. 
This cumulative value is designated as the positive part, while 
the remaining predictions are summed to form the negative part.
The CBR loss for f0 and note can be calculated as follow:
\begin{equation}
    {L_{f0}} = \frac{1}{B}\sum\limits_{i = 1}^B {CE(a_i^s,a_i^w)},
\label{eq}
\end{equation}
\begin{equation}
    {L_{note}} = \frac{1}{B}\sum\limits_{i = 1}^B {CE(b_i^s,b_i^w)},
\label{eq}
\end{equation}
where $a_i^s$ and $a_i^w$ represent the positive-negative probabilities of the f0
for strongly and weakly augmented data, respectively, 
while $b_i^s$ and $b_i^w$ denote the corresponding probabilities for the note.
Finally, the total CBR loss for unlabeled data is defined as:
\begin{equation}
    {L_u} = {L_{f0}} + {L_{note}}.
    \label{eq}
    \end{equation}

	\begin{table}[t]
		\caption{
			Results of Ablation Study on ADC2004, MIREX 05 and MedleyDB dataset. 
			The values in the table are percentile. ${\rm{SpectMamb}}{{\rm{a}}^N}$
			stands for the proposed model without  note-f0 decoder and
			${\rm{SpectMamb}}{{\rm{a}}^B}$ stands for the propose moedel without
			CBR module,
			respectively.
		}
		\begin{center}
			\begin{tabular}{|c|c|c|c|c|c|c|}
				\hline
				\multirow{2}{*}{\centering \textbf{Method}}&\multicolumn{3}{|c|}{\textbf{ADC2004}}&\multicolumn{3}{|c|}{\textbf{MIREX 05}} \\
				\cline{2-7} 
				 & \textbf{\textit{OA}}& \textbf{\textit{RPA}}&\textbf{\textit{RCA}}& \textbf{\textit{OA}}& \textbf{\textit{RPA}}& \textbf{\textit{RCA}}\\
				\hline
				\text{${\rm{SpectMamb}}{{\rm{a}}^N}$}&  \text{78.58}& \text{77.29}& \text{77.61}&\text{82.85}& \text{78.64}& \text{79.25}\\
				\hline
				\text{${\rm{SpectMamb}}{{\rm{a}}^B}$}& \text{79.22}& \text{78.32}&\text{78.66}&  \text{83.24}& \text{80.25}& \text{80.61}\\
				\hline
				
				\text{SpectMamba}&  \textbf{80.67}& \textbf{80.62}&\textbf{80.95}& \textbf{84.64}& \textbf{81.82}& \textbf{82.16}\\
				\hline
			\end{tabular}
			\label{tab1}
		\end{center}
	\end{table}

\section{EXPERIMENTS}
\subsection{Experiment Setup}
To simulate the limited labeled data scenario, 
we select 400 music tracks from MIR-1K \cite{b11} and 35 music tracks 
from MedleyDB \cite{b12} as the training data. Then we also choose 1000 
music tracks from FMA dataset \cite{b13} as unlabeled data. 
For the testing data, we use three well-known testing datasets: 
12 tracks from ADC2004, 9 tracks from 
MIREX 05\footnote{https://labrosa.ee.columbia.edu/projects/melody}, 
12 tracks from MedleyDB. 
In accordance with the practice in the 
literature \cite{b14}, the performance evaluation is based on the following 
metrics: voicing recall (VR), and voicing false alarm (VFA), raw f0 
accuracy (RPA), raw chroma accuracy (RCA) and overall accuracy (OA). 
The mir\_eval \cite{b15} library is employed to compute these metrics. 
The performance is measured by the score of each metric except VFA, 
where a higher score indicates better performance. OA is 
regarded as a more important metric 
than the others 
in the literature \cite{b16}.

\subsection{Implementation Details}
To prepare the input data to our proposed model, 
following \cite{b14} we use CFP \cite{b22,b23,b24,b25} as input representations.
The raw waveform is resampled to 8,000 Hz, 
window size is set to 768, and hop size is set to 80 for 
computing the short-time-Fouriertransformation (STFT). When 
computing CFP, we use specific setting as follow: 60 bins 
per octave, with the number of frequency bins to 320, and a 
frequency range from 31 Hz (B0) to 1250 Hz (${\rm{D\# 6}}$). Our 
model is implemented with PyTorch\footnote{https://pytorch.org}. The Adam optimizer 
is used with the learning rate of 0.0004.
$\omega$ in loss function is set to 0.1.

\begin{table}[t]
	\caption{The experimental results of SpectMamba and baseline models on ADC2004, MIREX 05 and MedleyDB datasets.}
	\begin{center}
		\begin{tabular}{|c|c|c|c|c|c|}
			\multicolumn{6}{c}{\textbf{(a) ADC2004}}\\
			\hline
			\multirow{2}{*}{\centering \textbf{Method}}&\multicolumn{5}{|c|}{\textbf{ADC2004}} \\
			\cline{2-6} 
			 &  \textbf{\textit{VR}}& \textbf{\textit{VFA}}&\textbf{\textit{OA}}& \textbf{\textit{RPA}}& \textbf{\textit{RCA}}\\
			\hline
			\text{FTANet}&  \text{79.62}& \textbf{5.65}& \text{77.15}&\text{74.95}& \text{74.98}\\
			\hline
			\text{TONet}&  \text{83.62}& \text{16.84}&\text{78.34}& \text{75.59}& \text{75.97} \\
			\hline
			\text{${{\rm{S}}^{\rm{2}}}{\rm{Former}}$}&  \text{84.47}& \text{17.53}&\text{78.43}& \text{78.70}& \text{79.13} \\
			\hline
			\text{SpectMamba(Ours)}&  \textbf{85.82}& \text{17.51}&\textbf{80.67}& \textbf{80.62}& \textbf{80.95}\\
			\hline
			
		\end{tabular}
		\label{tab1}
	\end{center}
	\begin{center}
		\begin{tabular}{|c|c|c|c|c|c|}
			\multicolumn{6}{c}{\textbf{(b) MIREX 05}}\\
			\hline
			\multirow{2}{*}{\centering \textbf{Method}}&\multicolumn{5}{|c|}{\textbf{MIREX 05}} \\
			\cline{2-6} 
			&  \textbf{\textit{VR}}& \textbf{\textit{VFA}}& \textbf{\textit{OA}}&\textbf{\textit{RPA}}& \textbf{\textit{RCA}}\\
			\hline
			\text{FTANet}& \text{77.80}& \textbf{4.19}&\text{82.91}&  \text{75.13}& \text{75.21} \\
			\hline
			\text{TONet}&  \text{88.02}& \text{13.49}&\text{83.20}& \text{80.89}& \text{81.27} \\
			\hline
			\text{${{\rm{S}}^{\rm{2}}}{\rm{Former}}$}&  \text{82.96}& \text{7.90}&\text{82.79}& \text{77.53}& \text{77.61}\\
			\hline
			\text{SpectMamba(Ours)}& \textbf{88.26}& \text{11.38}& \textbf{84.64}&  \textbf{81.82}& \textbf{82.16}\\
			\hline
		\end{tabular}
		\label{tab1}
	\end{center}
	\begin{center}
		\begin{tabular}{|c|c|c|c|c|c|}
			\multicolumn{6}{c}{\textbf{(c) MedleyDB}}\\
			\hline
			\multirow{2}{*}{\centering \textbf{Method}}&\multicolumn{5}{|c|}{\textbf{MedleyDB}} \\
			\cline{2-6} 
			&  \textbf{\textit{VR}}& \textbf{\textit{VFA}}&\textbf{\textit{OA}}& \textbf{\textit{RPA}}& \textbf{\textit{RCA}}\\
			\hline
			\text{FTANet}&  \text{55.62}& \text{9.87}&\text{69.28}& \text{49.85}& \text{50.90} \\
			\hline
			\text{TONet}&  \text{61.99}& \text{13.94}&\text{69.46}& \text{53.75}& \text{55.14} \\
			\hline
			\text{${{\rm{S}}^{\rm{2}}}{\rm{Former}}$}&  \text{58.96}& \textbf{9.45}&\text{71.35}& \text{53.39}& \text{54.28} \\
			\hline
			\text{SpectMamba(Ours)}&  \textbf{64.61}& \text{12.66}& \textbf{72.62}&\textbf{58.14}& \textbf{58.96} \\
			\hline
			
		\end{tabular}
		\label{tab1}
	\end{center}
\end{table}

\subsection{Ablation Study}
An ablation study is conducted to investigate the impact of each module 
to our proposed model. As shown 
in Table 1, we first remove the note-f0 decoder and
the performance on two datasets is decreased. When 
focusing on OA, the performance is decreased by 2.09\% on ADC2004, and
1.79\% on MIREX 05.
The observation indicates that the use of note supervision
helps improve the performance of singing melody extraction.
We then remove the
CBR module and keep the note-f0 decoder. 
The performance on two datasets is decreased by 
1.45\% on ADC2004, and 1.4\% on MIREX 05. 
The results show that the proposed 
modules are tightly cooperated and contribute to the proposed SpectMamba.

\subsection{Comparison with State-of-the-art Models}
The experimental results on the three public datasets 
are shown in Table 2. Three baseline models are compared 
in Table 2, including FTANet \cite{b2}, TONet \cite{b19} and 
${{\rm{S}}^{\rm{2}}}{\rm{Former}}$ \cite{b18}. 
The proposed model and the three baseline models are 
\begin{figure}[b]
    \centerline{\includegraphics[scale=0.30]{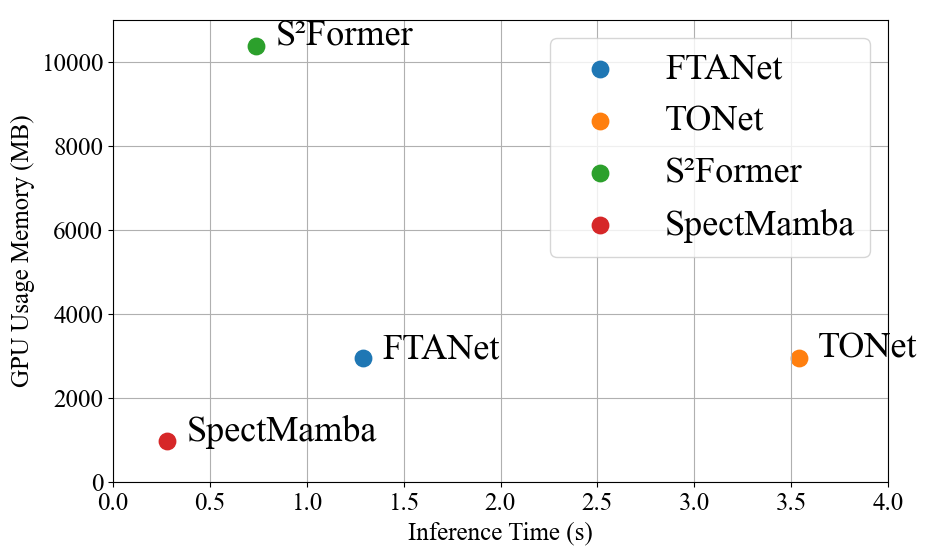}}
    \caption{Inference Time vs GPU Usage Memory in ADC2004.}
    \label{fig}
    \end{figure}

trained 
on the same training set. Compared with the three baseline 
models, the proposed model achieves the highest results in 
general. 
Compared with transformer-based model 
on OA, our proposed model outperforms ${{\rm{S}}^{\rm{2}}}{\rm{Former}}$
2.24\% in ADC2004, by 1.85\% in MIREX 05, and by 1.27\% 
in Medley DB. It is worth mentioning that the 
proposed model demonstrates better performance 
than traditional transformers.

\begin{figure}[t]
    \centering
    \begin{minipage}[b]{0.241\textwidth} 
        \centering
        \includegraphics[width=\textwidth]{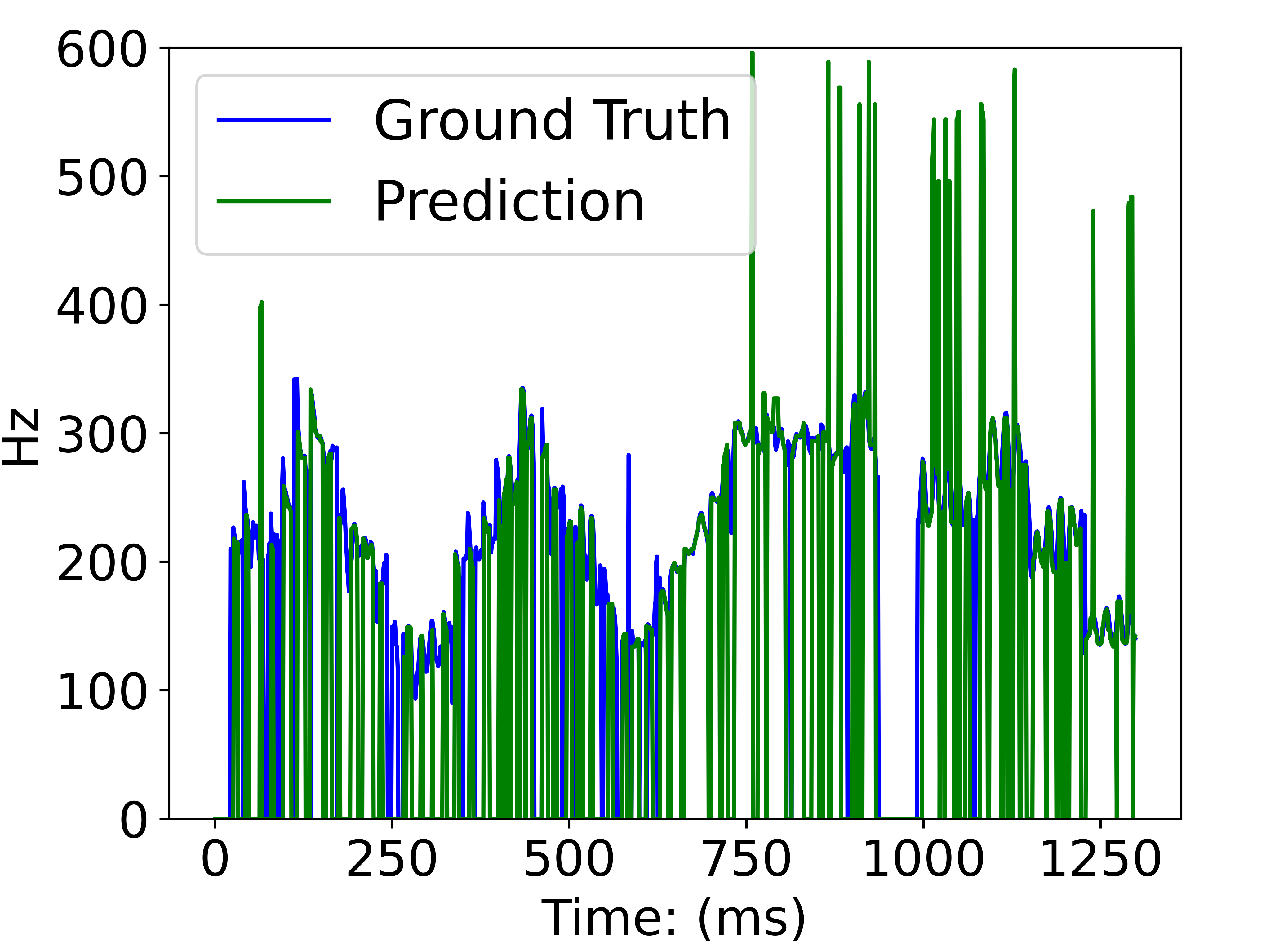} 
        {\scriptsize (a) Opera\_male3 by ${{\rm{S}}^{\rm{2}}}{\rm{Former}}$ \cite{b18}}
    \end{minipage}
    \begin{minipage}[b]{0.241\textwidth} 
        \centering
        \includegraphics[width=\textwidth]{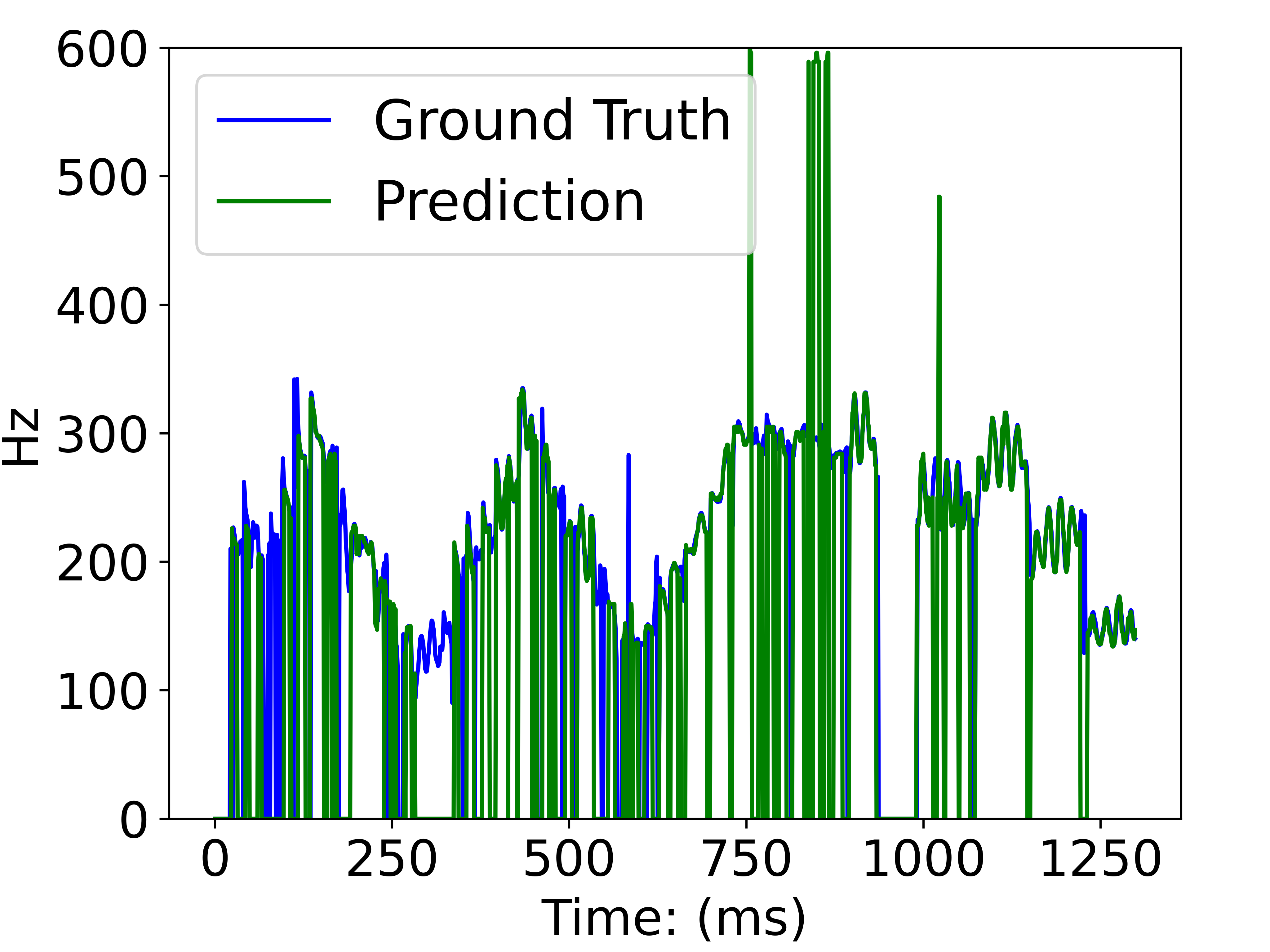}
        {\scriptsize (b) Opera\_male3 by our model.}
    \end{minipage}

    \begin{minipage}[b]{0.241\textwidth}
        \centering
        \includegraphics[width=\textwidth]{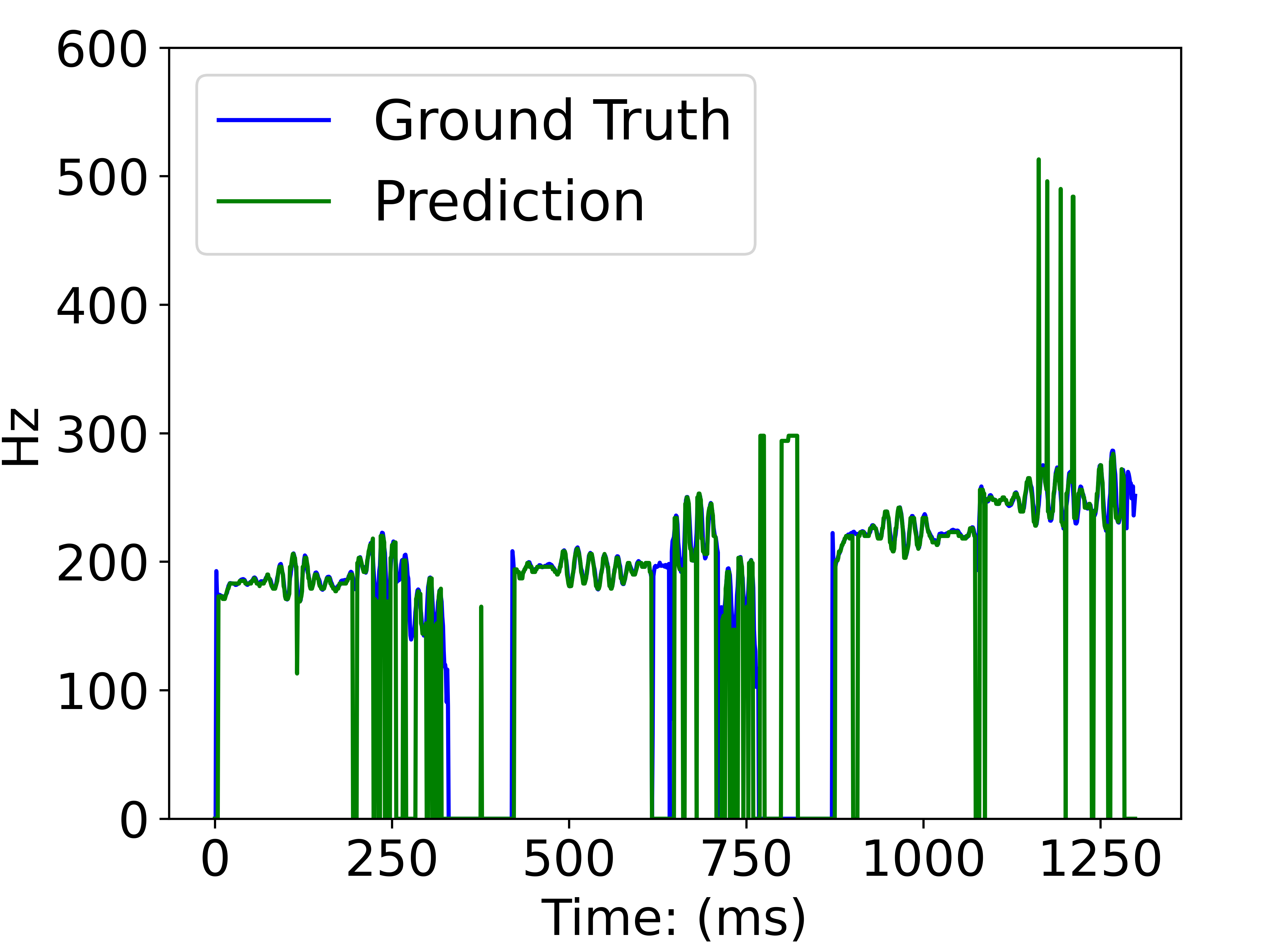}
        {\scriptsize (c) Opera\_male5 by ${{\rm{S}}^{\rm{2}}}{\rm{Former}}$ \cite{b18}}
    \end{minipage}
    \begin{minipage}[b]{0.241\textwidth}
        \centering
        \includegraphics[width=\textwidth]{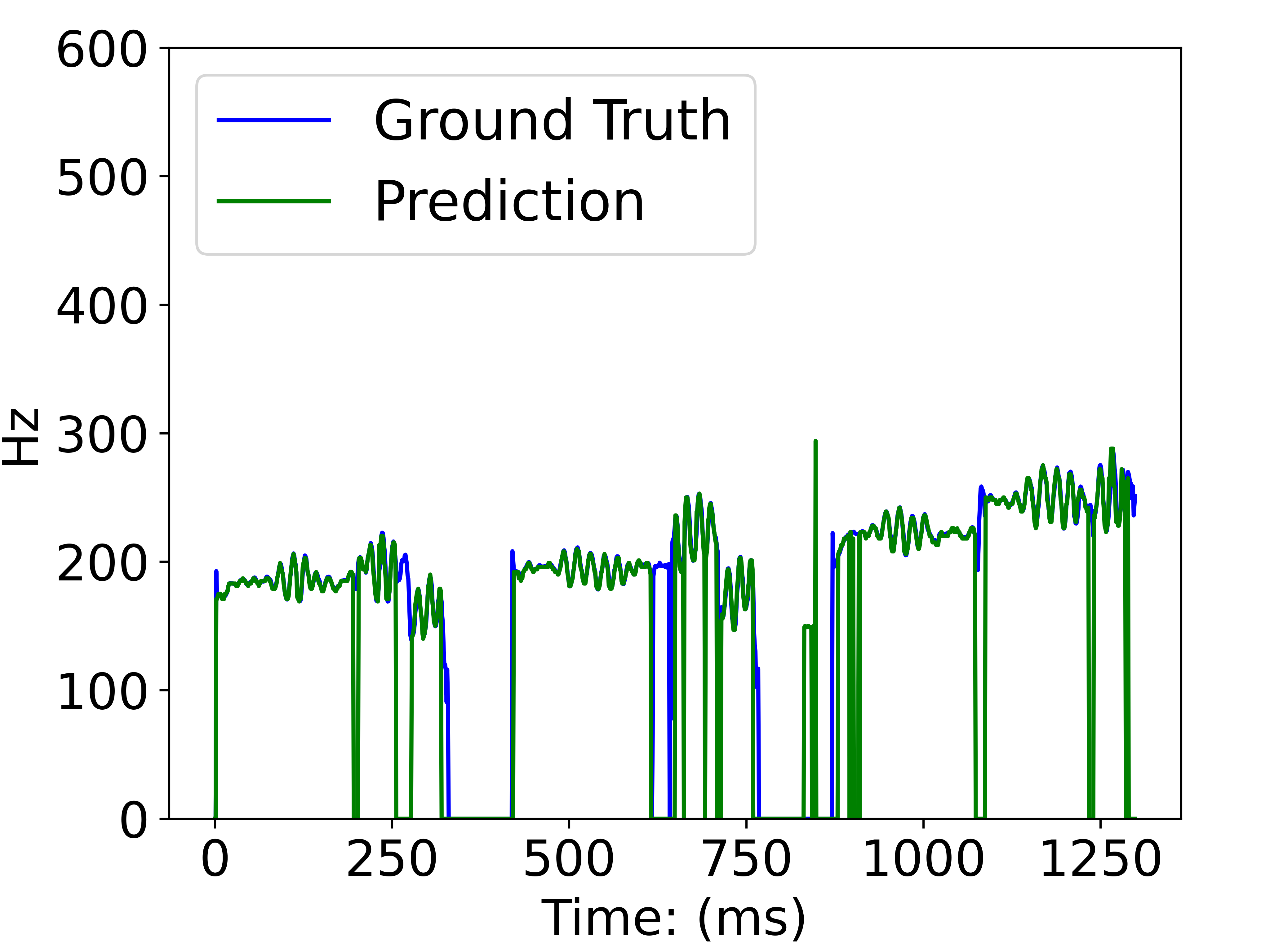}
        {\scriptsize (d) Opera\_male5 by our model.}
    \end{minipage}

    \caption{Visualization of singing melody extraction 
    results on two opera songs using ${{\rm{S}}^{\rm{2}}}{\rm{Former}}$ and the proposed SpectMamba.}
    \label{fig}
\end{figure}

We also compare GPU memory usage and inference time of the proposed
model with three baseline methods in ADC2004, and the results are shown in
Figure 2. We can observe that the proposed model outperforms 
the three baseline models in both inference time and GPU memory usage.
Compared with ${{\rm{S}}^{\rm{2}}}{\rm{Former}}$ which is based on transformer,
we achieve $2 \times $ the inference speed and only $1/10$ of the memory usage.

To further investigate how the proposed model improves the performance,
we have generated visualizations of the predictions for
two songs of opera singing in ADC2004. As shown in Figure 3,
the number of octave errors is notably lower in diagrams (b) and (d)
compared to (a) and (c). The visualization of predicted melody contour
demonstrates the effectiveness of our model.

\section*{CONCLUSION}
In this paper, we have proposed a mamba-based semi-supervised network, 
using both f0 and note supervision, along with the CBR module.
The mamba encoder improves the efficiency of the model;
the note-f0 decoder leverages musical prior knowledge to simulate music internal characteristics;
and the CBR module allows the model to effectively utilize unlabeled data. 
The proposed model is evaluated on a set of well-known public
melody extraction datasets with promising performances.
The experimental results demonstrate the effectiveness of the proposed network for 
singing melody extraction from polyphonic music.

\section*{ACKNOWLEDGMENT}
This work was supported by the Fundamental Research Funds
for the Central Universities (2232024D-27), NSFC (62171138).


\begin{thebibliography}{00}
\bibitem{b30} S Yu, X He, K Chen, and Y Yu, ``HKDSME: Heterogeneous Knowledge Distillation for Semi-supervised Singing Melody Extraction Using Harmonic Supervision," Proceedings of the 32nd ACM International Conference on Multimedia. 2024: 545-553.
\bibitem{b31} S Yu, X He, and Y Zhang, ``RevNet: A Review Network with Group Aggregation Fusion for Singing Melody Extraction," in Proc. ICME, 2024: 1-6.
\bibitem{b32} S Yu, ``MCSSME: Multi-Task Contrastive Learning for Semi-supervised Singing Melody Extraction from Polyphonic Music," Proceedings of the AAAI Conference on Artificial Intelligence. 2024, 38(1): 365-373.
\bibitem{b1} J Serra, E Gómez, and P Herrera, ``Audio cover song identification and similarity: background, approaches, evaluation, and beyond," in Proc. Advances in Music Information Retrieval, 2010, Springer, 307–332.
\bibitem{b2} S Yu, X Sun, Y Yu, and W Li, ``Frequency temporal attention network for singing melody extraction," in Proc. ICASSP, 2021, pp. 251–255.
\bibitem{b3} C C Wang and J S R Jang, ``Improving query-by-singing/humming by combining melody and lyric information," IEEE/ACM Trans. Audio Speech Language Processing, vol. 23, no. 4, pp. 798–806, 2015.
\bibitem{b4} S Yu, C Li, F Deng, and X Wang, ``Rethinking Singing Voice Separation With Spectral-Temporal Transformer," in Proc. APSIPA ASC, IEEE, 2021: 884-889.
\bibitem{b5} P Knees and M Schedl, ``Music retrieval and recommendation: A tutorial overview," in Proc. SIGIR, 2015, pp. 1133–1136.
\bibitem{b7} L Zhu, B Liao, Q Zhang, X Wang, W Liu, and X Wang, ``Vision mamba: Efficient visual representation learning with bidirectional state space model," arXiv preprint arXiv:2401.09417, 2024.
\bibitem{b11} C L Hsu and J S R Jang, ``On the improvement of singing voice separation for monaural recordings using the MIR-1K dataset," IEEE Trans. Speech Audio Process, vol. 18, no. 2, pp. 310–319, 2010.
\bibitem{b12} R M Bittner, J Salamon, M Tierney, M Mauch, C Cannam, and J P Bello, ``Medleydb: A multitrack dataset for annotation-intensive MIR research," in Proc. ISMIR, 2014, pp. 155–160.
\bibitem{b13} M Defferrard, K Benzi, P Vandergheynst, and X Bresson, ``FMA: A Dataset for Music Analysis," in Proc. 2017, ISMIR. 316–323.
\bibitem{b14} T H Hsieh, L Su, and Y H Yang, ``A streamlined encoder/decoder architecture for melody extraction," in Proc. ICASSP, 2019, pp. 156–160.
\bibitem{b15} C Raffel, B McFee, E J Humphrey, J Salamon, O Nieto, D Liang, D P Ellis, and C C Raffel, ``mir\_eval: A transparent implementation of common mir metrics," in Proc. ISMIR, 2014.
\bibitem{b16} J Salamon, E Gómez, D P W Ellis, and G Richard, ``Melody extraction from polyphonic music signals: Approaches, applications, and challenges," IEEE Signal Processing Magazine, vol. 31, no. 2, pp. 118–134, 2014.
\bibitem{b17} W T Lu, J C Wang, M Won, K Choi, and X Song, ``SpecTNT: a time-frequency transformer for music audio,” in Proc. ISMIR, 2021, pp. 396–403.
\bibitem{b18} S Yu, J Liu, Y Yu, and W Li, ``A scalable sarse Transformer model for singing melody extraction," in Proc. ICASSP, 2024, pp. 1071-1075.
\bibitem{b19} K Chen, S Yu, C Wang, W Li, T Berg-Kirkpatrick, and Dubnov S, ``TONet: Tone-Octave Network for Singing Melody Extraction from Polyphonic Music," in Proc. ICASSP, 2022, 621–625.
\bibitem{a1} S Yu, Y Yu, X Chen, and W Li, ``HANME: hierarchical attention network for singing melody extraction," IEEE Signal Processing Letters, 2021, 28: 1006-1010.
\bibitem{a2} P Gao, C Y You, and T S Chi, ``A multidilation and multi-resolution fully convolutional network for singing melody extraction,” in Proc. ICASSP, 2020, 551–555.
\bibitem{a3} D Basaran, S Essid, and G Peeters, ``Main melody extraction with source-filter NMF and CRNN," in Proc. ISMIR, 2018.
\bibitem{a4} L Su, ``Vocal melody extraction using patch-based cnn,” in Proc. ICASSP, 2018, 371–375.
\bibitem{a5} S Yu, X Chen, and W Li, ``Hierarchical graph-based neural network for singing melody extraction," in Proc. ICASSP, 2022, 626–630.
\bibitem{silu} P Ramachandran, B Zoph, and Q V Le, ``Searching for activation functions," arXiv preprint arXiv:1710.05941, 2017.
\bibitem{b22} L Su and Y H Yang, ``Combining spectral and temporal representations for multipitch estimation of polyphonic music," J. IEEE/ACM Transactions on Audio, Speech, and Language Processing, 2015, 23(10): 1600-1612.
\bibitem{b23} L Su, T Y Chuang, and Y H Yang, ``Exploiting Frequency, Periodicity and Harmonicity Using Advanced Time-Frequency Concentration Techniques for Multipitch Estimation of Choir and Symphony," in Proc. ISMIR, 2016: 393-399.
\bibitem{b24} L Su, ``Between homomorphic signal processing and deep neural networks: Constructing deep algorithms for polyphonic music transcription," in Proc. APSIPA ASC. IEEE, 2017: 884-891.
\bibitem{b25} Y T Wu, B Chen, and L Su, ``Automatic music transcription leveraging generalized cepstral features and deep learning," in Proc. ICASSP, 2018, 401-405.
\bibitem{b26} S Kum, C Oh, and J Nam, ``Melody extraction on vocal segments using multicolumn deep neural networks,” in Proc. ISMIR, 2016.
\bibitem{b27} R M Bittner, B McFee, J Salamon, P Li, and J P Bello, ``Deep Salience Representations for F0 Estimation in Polyphonic Music," in Proc. ISMIR, 2017, 63–70.
\bibitem{b28} M T Chen, B J Li, and T S Chi, ``CNN Based Two-stage Multiresolution End-to-end Model for Singing Melody Extraction," in Proc. ICASSP, 2019, 1005–1009.
\bibitem{b29} S Yu, Y Yu, X Sun, and W Li, ``A neural harmonic-aware network with gated attentive fusion for singing melody extraction," Neurocomputing, 2023, 521: 160-171.
\bibitem{c30} Z Wu and J Cui, “AllMatch: Exploiting All Unlabeled Data for Semi-Supervised Learning,” arXiv preprint arXiv:2406.15763, 2024.
\end{thebibliography}
\end{document}